\def\real{{\tt I\kern-.2em{R}}}
\def\nat{{\tt I\kern-.2em{N}}}

\def\realp#1{{\tt I\kern-.2em{R}}^#1}
\def\natp#1{{\tt I\kern-.2em{N}}^#1}
\def\hyper#1{\ ^*\kern-.2em{#1}}

\def\St#1{{\tt st}#1}
\def\st#1{{\tt st}(#1)}
\def\hyperreal{{^*{\real}}}
\def\hyperrealp#1{{\tt ^*{I\kern-.2em{R}}}^#1} 
\def\hypernat{{^*{\nat }}}
\def\hypernatp#1{{{^*{{\tt I\kern-.2em{N}}}}}^#1} 
\def\eskip{\hskip.25em\relax}

\def\Hyper#1{\hyper {\eskip #1}}
\def\leaderfill{\leaders\hbox to 1em{\hss.\hss}\hfill}
\def\srealp#1{{\rm I\kern-.2em{R}}^#1}

\def\pars{\par\smallskip}
\def\parm{\par\medskip}

\def\ref#1{$^{#1}$}

\def\m@th{\mathsurround=0pt}
\def\rightarrowfill{$\m@th \mathord- \mkern-6mu \cleaders\hbox{$\mkern-2mu 
\mathord- \mkern-2mu$}\hfil \mkern-6mu \mathord\rightarrow$}
\def\leftarrowfill{$\mathord\leftarrow
\mkern -6mu \m@th \mathord- \mkern-6mu \cleaders\hbox{$\mkern-2mu 
\mathord- \mkern-2mu$}\hfil $}
\def\noarrowfill{$\m@th \mathord- \mkern-6mu \cleaders\hbox{$\mkern-2mu 
\mathord- \mkern-2mu$}\hfil$}
\def\orgate{$\bigcirc \kern-.80em \lor$}
\def\andgate{$\bigcirc \kern-.80em \land$}
\def\inverter{$\bigcirc \kern-.80em \neg$}
\magnification=\magstep1  
\tolerance 10000
\baselineskip  13pt
\hoffset=0.25in
\hsize 6.00 true in
\vsize 8.75 true in

\def\id{\par\hangindent2\parindent\textindent}
\def\textindent#1{\indent\llap{#1}}  
\centerline{{\bf The Encoding of Quantum State Information Within Subparticles}}\bigskip
\centerline{Robert A. Herrmann}\parm
\centerline{Mathematics Department}
\centerline{U. S. Naval Academy}
\centerline{572C Holloway Rd}
\centerline{Annapolis$,$ MD 21402-5002 USA}
\centerline{25 SEP 1999, last revision 22 NOV 2010.}\bigskip\par
{\leftskip = 0.5in \rightskip = 0.5in \noindent {\it Abstract:} A method is given by which the descriptive content of quantum state specific-information can be encoded into subparticle coordinates. This method is consistent with the GGU-model solution to general grand unification problem.\par}\par\bigskip
\noindent {\bf 1. State specific-information.}\par\medskip  
Subparticles are predicted by the GGU-model (Herrmann$,$ 1993, 1994). They are used for various purposes and it has recently been shown that subparticles can transmit quantum state specific-information over any distance in a manner that would appear to be instantaneous (Herrmann$,$ 1999a) without violating Einstein separability. 
What has been missing from these investigations is a direct method that would encoded such specific-information within the appropriate ultrasubparticle coordinate so that the generation of an intermediate subparticle would display this specific-information when the standard part operator$,$ $\st {\cdot},$ is applied. \pars 
Recall that each of the coordinates $i\geq 3$ of an ultrasubparticle that characterizes physical behavior contains only the numbers $\pm 1/10^{\omega}\in \mu(0),$
where $\omega$ is a Robinson (1966) infinite natural number and $\mu(0)$ the set of hyperreal infinitesimals. 
Consider any descriptive scientific language $\cal L,$ and construct an intuitive word $w$ that yields the descriptive content for a particular quantum state. Note that $w\in [w]\subset {\cal L},$ where $[w]$ is the equivalence class of all symbol strings that have the same relative meaning
(Herrmann$,$ 1999b). Then$,$ as in Herrmann (1987, 1993, 1994)$,$ there is an encoding $i$ into the set of natural numbers (including 0). In the equivalence classes of partial sequences $\cal E,$ there is the unique set $[f_0]$ such that $f_0(0)= i(w).$ \pars 

From subparticle theory (Herrmann$,$ 1993, 1994$,$ 1999a)$,$ there is an ultrasubparticle coordinate $a_j$ that corresponds to this quantum state specific-information. It is also known (Herrmann$,$ 1993, 1994$,$ 1999a)$,$ that there exists a hyperreal natural number $\lambda$ such that $\st {\lambda/10^\omega} = i(w).$ 
Hence, in order to obtain the intermediate subparticle such that $\st {a_j} = i(w),$ simply consider a $\lambda$ hyperfinite combination of  ultrasubparticles. This will yield an intermediate subparticle with an $a_j$ coordinate 
$$\st {\sum_{n = 1}^\lambda {{1}\over{10^\omega}}} = i(w),\eqno(1.1)$$
where each term of this hyperfinite series is the constant $1/10^\omega.$
Since $i$ is an injection$,$ the actual state specific-information$,$ in descriptive form$,$ 
can be re-captured by considering the inverse $i^{-1}.$ What this implies is that by hyperfinite subparticle combination and application of the standard part operator each elementary physical particle may be assumed to be composed of all of the requisite state specific-information that distinguishes it from any other quantum physical entity and state specific-information is characterized by a specific coordinate entry. \parm

\noindent {\bf 2. Subparticle mechanisms.}\par\medskip 

The mapping $i,$ in section 1, is somewhat arbitrary. The ``word'' $w$ merely represents the specific-information it contains. As stated in Herrmann (1993) about subparticles, ``These multifaceted things, these subparticles are not to be construed as either particles nor waves nor quanta nor anything that can be represented by a fixed imagery. Subparticles are to be viewed only operationally'' (p. 99). The hyperfinite sum (1.1) is an analogue model for an {\it ultrafinite combination of ultrasubparticles}. 
Each ultrafinite combination can be conceived of as a ``bundling'' of the ultrasubparticles. An intermediate subparticle is generated by the ultralogic $\Hyper {\bf S}$ applied to an ultraword called an {\it ultramixture}. Each elementary particle is generated by $\Hyper {\bf S}$ applied to a single ultraword termed an {\it ultimate ultramixture} (Herrmann, 1993, p. 106). This method of generation can be continued through all levels of complexity. Of course, realization occurs only when the standard part operator is applied. [Note: When the physical-like process is discussed the term ``ultrafinite'' is used rather than ``hyperfinite'' as an indication that this applies to a specific physical-like process and a mathematical model such as $\st {\sum_{n = 1}^\lambda {{1}\over{10^\omega}}} = i(w)$.]\pars 
  
The infinitesimal values $\pm 1/10^\omega$ used in ultrasubparticle coordinates $a_i,\ i \geq 3$ are, in general, not unique. Thus, mathematical expressions that appear in this section only represent analogue models for subparticle behavior. This is especially the case where defined measures are used as representations for the ``qualities'' that characterize the elementary ``particles'' since the values of the various $\lambda \in \nat_\infty$ used in hyperfinite sums such as (1.1) are dependent upon a defined measure. Hence, they but {\it represent} the notion of an ultrafinite combination of ultrasubparticles. \pars

All that has been done with subparticle theory can be accomplished using other forms such as $\pm 1/2^\omega.$ Everything in subparticle theory, including the next material, can be done in the binary form.  The actual mathematical model itself states that there is much specific behavior within the model that cannot be expressed in any human language form. This does not eliminate general descriptions for behavior, descriptions that need not refer to specific quantities as a being unique. Such general descriptions follow the same patterns as the descriptions that appear in the first paragraph of this section. 
\pars 
In Herrmann (1993, p. 103), {\it independent *-finite coordinate summation} is used to model the formation of an {intermediate subparticle} generated by $\Hyper {\bf S}$ applied to an ultramixture. It is mentioned that this can also be accomplished operationally by repeated application of an affine (actually a *-affine) transformation. In order to discuss various transformations that model subparticle mechanisms, objects equivalent to matrices are used. These matrices are viewed in our Nonstandard Model as internal functions $M\colon [1,n]\times [1,n]\to \hyperreal,$ where nonzero $n \in \hypernat.$ It is not difficult to convent, by *-transform, the usual process of multiplying a standard $(a_1,\ldots, a_n)$ by a standard $n \times n$ matrix $\rm A$ and obtain a transformed standard $(b_1,\ldots, b_n)$ into statements using $M$. The result obtained will be a hyperfinite and, hence, internal object. Further, the usual linear algebra finite coordinatewise summation *-transferred to the hyperfinite yields internal objects. When appropriate, the usual matrix notation is used.\pars 
In Herrmann (1993, p. 110), the possible ``naming'' of ultrafinite combinations is discussed. [There is a typographical error in (2), the last line on this page. It should read ``For the set of all fundamental entities. . . . entities.''] This process does not lend itself to a simple *-affine transformation. However, under the proper specific-information interpretation, the first and second coordinates yield no natural-world specific-information and, when realization occurs, are suppressed. {\it Indeed, for only natural-world applications one need not include these two coordinates as part of the basic operational definition for subparticles and if not included, what follows would need to be slightly modified.} Consequently, the following approach should be considered as but models for the production of natural-world specific-information. \pars

In what follows, the most general ultrasubparticle case is used, where $a_1 = k,\ a_2 = 1.$ For a single characteristic modeled by coordinate $a_i,\ i > 2$, one of the simplest transformation processes is modeled by the internal *-affine transformation $${\rm T}(\cdot)^{n\times 1}= {\rm I}(\cdot)^{n\times 1}+ (b_j)^{n \times 1}, \eqno (2.1)$$
where $\rm I$ is an identity matrix, $b_1=0,\ b_2 = 1,\ b_i =a_i$ and $b_j = 0,\ j \not=i, 3\leq j \leq n.$ 
This transformation must be applied $\lambda -1$ times. For example, assume that $\rm T$ is ``first'' applied to $(k,1,1/10^\omega,-1/10^\omega,\ldots,a_n)^T,\ \omega \in \nat_\infty,\ n \in \hypernat$ and models a quality expressed by coordinate $i = 3$. Then 
applying the *-translation $\rm T$ for a total of ``$\lambda -1$ times'' yields the intermediate subparticle $$(k,\lambda,\sum_{n=1}^\lambda 1/10^\omega,-1/10^\omega,\ldots,a_n)=(k,\lambda,\lambda/10^\omega,-1/10^\omega,\ldots,a_n).\eqno (2.2)$$ 
In this form, the second coordinate reveals the count value. \pars

The operator that yields the realized intermediate subparticle for (2.2) can be represented by a matrix composed of operators.  This matrix is $(b_{ij})^{n\times n},$ where $b_{ii} = \St,\ 3\leq i\leq n$ and $0 =b_{11}=b_{22} = b_{ij},\ i \not= j,\ 1\leq i \leq n,\ 1 \leq j \leq n$ and, for this case, yields $(0,0,r,0,\ldots,0), \ \st {\lambda/10^\omega} = r,$ a representation for specific-information. This matrix can be modified relative to the first two coordinates since these coordinates do not yield any natural-world specific-information.  There are relations between coordinate specific-information and it is these relations that determine how the elementary particles behave when they are conceived of as finite collections of intermediate subparticles. \pars
Rather than (2.1), there are more specific $\lambda$ dependent internal *-affine transformations which need apply but once as a model for coordinate independent *-finite coordinate summation. For the $\lambda$, simply consider 
$${\rm T}(\lambda)(\cdot)^{n\times 1}= {\rm I}(\cdot)^{n\times 1}+ (b_j)^{n \times 1}, \eqno (2.3)$$
where $b_1=0,\ b_2 = \lambda - 1,\ b_i =(\lambda -1)a_i$ and $b_j = 0,\ j \not= i, 3\leq j \leq n.$ 
One application of (2.3) yields (2.2).\pars

The generation of the intermediate subparticles by these methods and then the bundling of finitely many to yield any elementary particle in any state can be represented by a single and simple internal *-linear transformation upon which the standard part operator is applied. Suppose that there is but three qualities necessary to completely describe an elementary particle and these are modeled by $\lambda_1,\lambda_2,\lambda_3$ and coordinates 3,4,5. Consider matrix $(b_{ij})^{n \times n},\ b_{ii}=0;\ i=1,2,6,7,\ldots,\ b_{33} =\lambda_1,\ b_{44} =\lambda_2,\ b_{55} = \lambda_3$ and $b_{ij} = 0, i \not=j, 1\leq i\leq n,\ 1\leq j \leq n.$ For realization, consider $\st {(b_{ij})(k,1,1/10^\omega,-1/10^\omega,\ldots,a_n)^T} = (0,0,r_1,r_2,r_3,0,\ldots,0).$ If one wanted this matrix to replicate the finite bundling of intermediate subparticles with suppression of the naming and the count number coordinates rather than merely producing the standard results, then the $b_{ii}= 3,\ i\geq 6$ .\pars 
The major purpose for this section is to establish that the general notion of independent *-finite coordinate summation, technically, is not a forced   mathematical procedure designed solely for this one purpose since the results can be duplicated by other rather simple and well known mathematical operators. However, rigorously establishing the results by other well known means is less significant than the intuitive idea of generating specific-information from infinitesimal pieces of information by a simple bundling process, and then continuing this bundling process to obtain elementary particles. Retaining these intuitive notions remains the major purpose for (a) the independent *-finite summation model followed by (b) the usual coordinatewise summation for coordinates $i \geq 3$. The (b) process can be continued to obtain the specific-information that characterizes more complex entities composed of the assumed elementary particles. \parm   
                     
\centerline{\bf References}\parm 
\id{H}errmann$,$ R. A. 1987. Nonstandard consequence operators, Kobe  
J. Math., 4(1)(1987), 1-14.
\id{H}errmann, R. A. 1993. The Theory of Ultralogics Part II.\hfil\break http://www.arxiv.org/abd/math.GM/9903082     
\id{H}errmann$,$ R. A. 1994. Solution to the General Grand Unification Problem. 
http://www.arxiv.org/abs/astro-ph/9903110
\id{H}errmann$,$ R. A. 1999a. The NSP-World and Action-at-a-Distance. in  Vol. 2$,$  {\it Instantaneous Action-at-a-Distance in Modern Physics: ``Pro'' and ``Contra''} ed. Chubykalo$,$ A.$,$  N. V. Pope and R. Smirnov-Rueda$,$ Nova Science Books and Journals$,$ New York:223-235.
\id{H}errmann, R. A. 1999b. The Wondrous Design and Non-random Character of Chance Events. http://www.arxiv.org/abs/physics/9903038
\id{R}obinson, A. 1966. Non-standard analysis. North-Holland, Amsterdam.\par
\id{R}udin, W. 1964. The Principles of Mathematical Analysis, McGraw-Hill, New York.\par

\end